%% file: so_spl.tex
\documentclass[10pt,twocolumn,twoside]{IEEEtran}

\usepackage{color}     
\usepackage{epsf,psfrag,amssymb,amsfonts,latexsym,cite,verbatim,enumerate}
\usepackage{srcltx,amsmath,cases, graphicx}
\usepackage[mathscr]{eucal}
\usepackage{subfigure, stfloats}

\input macros.tex

\input labelfig.tex

\newcommand {\Ebb}{{\mathbb{E}}}

\def\Ebb{{\mathbb E}}

\newcommand{\beq}{\begin{equation}}
\newcommand{\eeq}{\end{equation}}

\newtheorem{remark}{Remark}

\newtheorem{problem}{Problem}
\newtheorem{proposition}{Proposition}

\title{\LARGE {Pilot Signal Design for Massive MIMO Systems: A Received Signal-To-Noise-Ratio-Based Approach}}
\author{
Jungho So, Donggun Kim, Yuni Lee, {\em Student~Members,~IEEE}, and Youngchul
Sung$^\dagger$\thanks{$^\dagger$Corresponding author}, {\em
Senior~Member,~IEEE}
\thanks{The authors are with  Dept. of Electrical Engineering,  KAIST, Daejeon 305-701, South
Korea. E-mail:\{jhso, dg.kim, yunilee, and ycsung\}@kaist.ac.kr. This research was supported by Basic Science Research Program through the National Research Foundation of Korea (NRF)
funded by the Ministry of Education (2013R1A1A2A10060852).}}

\begin{document}

\maketitle

\begin{abstract}

In this paper, the pilot signal design for massive MIMO systems to
maximize the training-based received signal-to-noise ratio (SNR)
is considered under two channel models: block Gauss-Markov and
block independent and identically distributed
 (i.i.d.) channel models. First, it is shown that under the block
Gauss-Markov channel model, the optimal pilot design problem
reduces to a semi-definite programming (SDP) problem, which can be
solved numerically by a standard convex optimization tool. Second,
under the block i.i.d. channel model, an optimal solution is
obtained in closed form. Numerical results show that the proposed
method yields noticeably better performance than other existing
pilot design methods in terms of received SNR.
\end{abstract}


\section{INTRODUCTION}
\label{sec:intro}

Efficient channel estimation is a crucial problem for
massive multiple-input multiple-output (MIMO) systems  \cite{Rusek&Persson&Lau&Larsson&Edfors&Tufvesson&Marzetta:13SPM} and there is active research going on in this area \cite{Marzetta:10WCOM,Rusek&Persson&Lau&Larsson&Edfors&Tufvesson&Marzetta:13SPM,Hoydis&Brink&Debbah:13JSAC,Shepard&Yu&Anand&Li&Marzetta&Yang&Zhong:12MobiCom}.
While much research is conducted on time-division duplexing (TDD) massive MIMO systems \cite{Marzetta:10WCOM,Rusek&Persson&Lau&Larsson&Edfors&Tufvesson&Marzetta:13SPM,Hoydis&Brink&Debbah:13JSAC,Shepard&Yu&Anand&Li&Marzetta&Yang&Zhong:12MobiCom}, recently some researchers considered the problem of efficient channel estimation and pilot signal design for
more challenging frequency-division duplexing (FDD) massive MIMO systems in which the number of channel
parameters to estimate may be much larger than the resource allocated
to training. To quickly acquire a reasonable channel estimate with limited training resources, the authors in \cite{Noh&Zoltowski&Sung&Love:14arXiv,Noh&Zoltowski&Sung&Love:14ICASSP,Choi&Love&Bidigare:14arXiv} exploited the channel's spatial and temporal correlation under the framework of  Kalman filtering with the state-space channel model.  In particular,  the authors  in \cite{Noh&Zoltowski&Sung&Love:14arXiv,Noh&Zoltowski&Sung&Love:14ICASSP} considered the pilot signal design under the state-space (i.e., Gauss-Markov) channel model to minimize the channel estimation error, and showed that the channel can be estimated efficiently by properly designing the pilot signal and exploiting the channel statistics.
However, minimizing the channel estimation error is not the ultimate metric of data communication.  Hence, in this paper, we consider the optimal pilot signal design under the framework of the state-space channel model to maximize the received SNR\footnote{
In the multiple-input single-output MISO case, the training-based capacity is a monotone increasing function of the training-based received SNR \cite{Hassibi&Hochwald:03IT}. A training approach based on received SNR was considered in the context of feedback in \cite{Love&Choi&Bidigare:13CISS,Choi&Love&Bidigare:14arXiv}.  The difference of this paper from  \cite{Love&Choi&Bidigare:13CISS,Choi&Love&Bidigare:14arXiv} is that we here obtained an optimal pilot signal under the state-space channel model based on the training-based received SNR defined in \cite{Hassibi&Hochwald:03IT}, which is different from the SNR definition used in \cite{Choi&Love&Bidigare:14arXiv}.} for data transmission, which is sometimes a final goal of data communication.

\textbf{Notation:} We will make use of
standard notational conventions. Vectors and matrices are written
in boldface with matrices in capitals. All vectors are column
vectors. For a matrix $\Abf$, $\Abf^T$, $\Abf^H$, $\Abf^{-1}$,
$\mbox{Tr}(\Abf)$, $\mbox{rank}(\Abf)$, $\lambda_i(\Abf)$, and
$\Abf(i,j)$ indicate the transpose, conjugate transpose, inverse,
trace, rank,  $i$-th largest eigenvalue, and  $(i,j)$-th element
of $\Abf$, respectively. $\Lc(\Abf)$ denotes the linear subspace
spanned by the columns of $\Abf$, and $\Lc^\perp(\Abf)$ is the
orthogonal complement of $\Lc(\Abf)$. For a random vector $\xbf$,
$\Ebb\{\xbf\}$ denotes the expectation of $\xbf$, and
$\xbf\sim\mathcal{CN}(\mubf,\Sigmabf)$ means that $\xbf$ is
circularly-symmetric complex Gaussian-distributed with mean
$\mubf$ and covariance matrix $\Sigmabf$. $\Ibf$ and $\Obf$ denote
an identity matrix and an all-zero matrix, respectively.

\section{System Model and Background}
\label{sec:systemmodel}

In this paper, we  consider the same massive MISO system as that considered in
\cite{Noh&Zoltowski&Sung&Love:13ASILOMAR,Noh&Zoltowski&Sung&Love:14arXiv,Choi&Love&Bidigare:14arXiv}. The transmitter has
 $N_t$ transmit antennas,  the receiver has a single receive antenna ($N_t\gg1$), and   each transmit-receive antenna pair has flat fading. Under this model the received signal $y_i$ at symbol time $i$ is given by
\begin{equation}  \label{eq:measurementModel}
y_i = \sbf_i^H\hbf^{(i)} + n_i,\quad i=1,2,\ldots,
\end{equation}
where  $\sbf_i$ is the $N_t\times1$ transmit signal vector at symbol time $i$, $\hbf^{(i)}$ is the $N_t\times 1$ channel vector at symbol time $i$, and $n_i$ is the additive Gaussian noise at symbol time $i$ from $n_i \stackrel{i.i.d.}{\sim}\Cc\Nc(0,\sigma^2)$ with the noise variance $\sigma^2$.
For the channel model, we assume the stationary\footnote{We assume that stationarity holds at least locally  \cite{Stein:87JSAC,Molish:book}. That is, the channel statistics vary much slowly than channel's fast fading.}  block Gauss-Markov  vector process \cite{Noh&Zoltowski&Sung&Love:14arXiv,Choi&Love&Bidigare:14arXiv}. That is, the channel vector is constant over one block and changes to a different state at the next block according to the following model:
\begin{equation}
 \hbf_{l+1} = a\hbf_l+\sqrt{1-a^2}\bbf_{l}, ~\hbf_0 \sim\Cc\Nc(\mathbf{0},\Rbf_\hbf),~ l=0,1,\ldots\hspace{-.4em}\label{eq:channel_evoluation}
\end{equation}
where $\hbf_l$ is the channel vector for the $l$-th block, $a\in [0,1]$ is the temporal fading coefficient,  and $\bbf_{l}\stackrel{i.i.d.}{\sim}\Cc\Nc(\mathbf{0},\Rbf_\hbf)$ is the innovation vector at the $l$-th block independent of $\{\hbf_0,\ldots,\hbf_{l}\}$.  We assume that one block consists of $T$ symbols: The first $T_t$ symbols are used for training and the following $T_d=T-T_t$ symbols are used for unknown data transmission. Thus, we have
$\hbf^{(i)} = \hbf_l$ for $i=lT+m, ~m=1,2,\cdots,T$.
It is easy to verify the assumed time-wise stationarity, i.e., $\Rbf_\hbf = \Ebb\{\hbf_0\hbf_0^H\}=\Ebb\{\hbf_1\hbf_1^H\}=\cdots$, for the considered channel parameter setup. $\Rbf_\hbf$ captures the spatial correlation of the channel and depends on the antenna geometry and the scattering environment \cite{Kotecha&Sayeed:04SP}. We assume that $a$ and $\Rbf_\hbf$ are known to the system. (Please see \cite{Noh&Zoltowski&Sung&Love:14arXiv} regarding this assumption.) Let $\Rbf_\hbf = \Ubf\Lambdabf\Ubf^H$ be the eigen-decomposition  of $\Rbf_\hbf$, where $\Ubf$ is a $N_t \times R_c$ matrix composed of orthonormal columns and the $R_c \times R_c$ matrix $\Lambdabf$ contains all the non-zero eigenvalues of $\Rbf_\hbf$.
Since all $\{\hbf_l,l=0,1,\ldots\}$ are contained in the same subspace $\Lc(\Ubf)$,
 we can model the $l$-th block channel as $\hbf_l = \Ubf  \gbf_l$
because of the assumed stationarity.  Then, the channel dynamic  \eqref{eq:channel_evoluation} can be rewritten in terms of $\gbf_l$ as
\begin{equation}
 \gbf_{l+1} = a\gbf_l+\sqrt{1-a^2}\ebf_{l}, ~\gbf_0 \sim\Cc\Nc(\mathbf{0},\Lambdabf), ~l=0,1,\cdots\hspace{-.4em}\label{eq:channel_evoluation2}
\end{equation}
with $\ebf_l \stackrel{i.i.d.}{\sim}
\Cc\Nc({\mathbf{0}},\Lambdabf)$. (This random vector process is
again a stationary process with $\Ebb\{\gbf_l\gbf_l^H\}=\Lambdabf$
for all $l$).

By stacking the symbol-wise received signal in \eqref{eq:measurementModel} corresponding to the training period of each block, we have
\begin{equation}  \label{eq:measurement22}
\ybf_l = \Sbf_l^H \hbf_l + \nbf_l,
\end{equation}
where $\ybf_l=[y_{lT+1},y_{lT+2},\cdots,y_{lT+T_t}]^T$, $\Sbf_l = [ \sbf_{lT+1},\cdots,\sbf_{lT+T_t}]$, and $\nbf_l=[n_{lT+1},n_{lT+2},\cdots,n_{lT+T_t}]^T$. The total power allocated to the training period of each block is given by
$\mbox{Tr}(\Sbf_l\Sbf_l^H)\le\rho T_t$, which means that each pilot symbol has power $\rho$ on average.  Since $\hbf_l \in \Lc(\Ubf)$, there is no loss in setting $\sbf_i = \Ubf \tilde{\sbf}_i$ because the signal power allocated to $\Lc^\perp(\Ubf)$ will simply be  lost without affecting the received signal $y_i$.   Hence, we have
\begin{equation}
\Sbf_l=[\Ubf\tilde{\sbf}_{lT+1},\cdots,\Ubf\tilde{\sbf}_{lT+T_t}]=\Ubf\tilde{\Sbf}_l,
\end{equation}
where $\tilde{\Sbf}_l$ is a $R_c \times T_t$ matrix and we assume $R_c \ge T_t$, i.e., the number of symbols contained in one channel coherence time is smaller than the channel rank as in typical massive MIMO systems.
Then, the measurement model \eqref{eq:measurement22} is rewritten as
\begin{equation}  \label{eq:reduced_mmmodel}
\ybf_l = (\Ubf\tilde{\Sbf}_l)^H (\Ubf \gbf_l) + \nbf_l=\tilde{\Sbf}_l^H  \gbf_l + \nbf_l,
\end{equation}
and the power constraint on $\tilde{\Sbf}_l$ is given by $\mbox{Tr}(\tilde{\Sbf}_l\tilde{\Sbf}_l^H)=\mbox{Tr}(\Sbf_l\Sbf_l^H)\le\rho T_t$.  Thus, the original state-space model  \eqref{eq:channel_evoluation} and \eqref{eq:measurement22} is equivalent to the new model  (\ref{eq:channel_evoluation2}) and (\ref{eq:reduced_mmmodel}) under the known stationary subspace condition $\hbf_l = \Ubf \gbf_l$.
Under the state-space model (\ref{eq:channel_evoluation2}) and (\ref{eq:reduced_mmmodel}),  the optimal minimum mean-square-error (MMSE) channel estimation is given by Kalman filtering \cite{Kailath:book}.
That is,  the MMSE estimate $\hat{\gbf}_{l|l}$ and its estimation error covariance matrix $\Pbf_{l|l}$ are updated as follows \cite{Kailath:book}:
\begin{eqnarray}
\Kbf_l             &=& \Pbf_{l|l-1}\tilde{\Sbf}_l(\sigma^2\Ibf+\tilde{\Sbf}_l^H\Pbf_{l|l-1}\tilde{\Sbf}_l)^{-1}, \nonumber \\
\hat{\gbf}_{l|l}   &=& \hat{\gbf}_{l|l-1}+\Kbf_l(\ybf_{l}-\tilde{\Sbf}_l^H\hat{\gbf}_{l|l-1}), \nonumber\\
\Pbf_{l|l}         &=& (\Ibf-\Kbf_l\tilde{\Sbf}_l^H)\Pbf_{l|l-1}, \nonumber  \\
\hat{\gbf}_{l|l-1} &=& a\hat{\gbf}_{l-1|l-1}, \nonumber\\
\Pbf_{l|l-1}       &=& a^2\Pbf_{l-1|l-1}+(1-a^2)\Lambdabf, \label{eq:KalmanUpdate}
\end{eqnarray}
where $\hat{\gbf}_{l|l^\prime} :=\Ebb\{\gbf_l|\ybf_0,\ybf_1,\cdots,\ybf_{l^\prime}\}$, $\Pbf_{l|l^\prime}:=\Ebb\{(\gbf_l-\hat{\gbf}_{l|l^\prime})(\gbf_l-\hat{\gbf}_{l|l^\prime})^H\}$,   $\hat{\gbf}_{0|-1}=\mathbf{0}$, and $\Pbf_{0|-1}=\Lambdabf$.

\section{Problem Formulation}
\label{sec:problem_formulation}

In this section, we consider the pilot design problem to maximize {\em the received SNR for  the data transmission period} under the assumption that $T$ and $T_t$ are given and the transmit beamforming  is used for the considered MISO channel during the data transmission period, i.e.,
\begin{equation}  \label{eq:beamformerweight}
\sbf_i =\wbf_i d_i=\Ubf\tilde{\wbf}_i d_i, ~~~i=lT+T_t+m,~~m=1,\cdots,T_d,
\end{equation}
where $\wbf_i$ and $d_i$ are the transmit beamforming vector and data symbol for symbol time $i$. Here, we assume $\Ebb\{d_i\}=0$ and $\Ebb\{|d_i|^2\}=\sigma_d^2$. From here on, we set $\sigma^2=1$ for simplicity. Again due to $\hbf_l \in \Lc(\Ubf)$, we can set $\wbf_i=\Ubf\tilde{\wbf}_i$ without any performance  loss.
From now on, we use $i(l)$ instead of $i$ for $i=lT+m,~m=1,\cdots,T$. First, following the framework in \cite{Hassibi&Hochwald:03IT}, we derive the received SNR during the data transmission period. The true channel at symbol time $i(l)$ is expressed as
\begin{equation}  \label{eq:channel_est_error}
\hbf_{l(i)} = \hat{\hbf}_{l(i)|l(i)} + \Delta \hbf_{l(i)},
\end{equation}
where $l(i)$ is the block number corresponding to symbol time $i(l)$, $\hat{\hbf}_{l(i)|l(i)}:=\Ubf\hat{\gbf}_{l(i)|l(i)}$ with $\hat{\gbf}_{l(i)|l(i)}$ obtained from \eqref{eq:KalmanUpdate} is the MMSE estimate for $\hbf_{l(i)}(=\Ubf\gbf_{l(i)})$  (this is true because $\mbox{Tr}(\Ebb\{(\gbf_l-\hat{\gbf}_{l|l})
(\gbf_l-\hat{\gbf}_{l|l})^H\})=\mbox{Tr}(\Ebb\{(\gbf_l-\hat{\gbf}_{l|l})
(\gbf_l-\hat{\gbf}_{l|l})^H\}\Ubf^H\Ubf)=\mbox{Tr}(\Ebb\{(\hbf_l-\hat{\hbf}_{l|l})(\hbf_l-\hat{\hbf}_{l|l})^H\})$), and $\Delta \hbf_{l(i)}$ is the channel estimation error.  Substituting \eqref{eq:beamformerweight} and \eqref{eq:channel_est_error} into \eqref{eq:measurementModel}, we have
\begin{align}
y_{i(l)} &= d_{i(l)}\wbf_{i(l)}^H(\hat{\hbf}_{l(i)|l(i)} + \Delta \hbf_{l(i)}) + n_{i(l)}, \nonumber\\
&= d_{i(l)}\tilde{\wbf}_i^H \hat{\gbf}_{l(i)|l(i)} + (d_{i(l)}\tilde{\wbf}_i^H\Delta \gbf_{l(i)} + n_{i(l)}). \label{eq:reduced_mmmodel2}
\end{align}
The key point in \cite{Hassibi&Hochwald:03IT} is that in the right-hand side (RHS) of  \eqref{eq:reduced_mmmodel2}, the term $\tilde{\wbf}_{i(l)}^H\hat{\gbf}_{l(i)|l(i)}$ is known to the receiver and the terms $\tilde{\wbf}_{i(l)}^H\Delta \gbf_{l(i)}$ and $n_{i(l)}$ are unknown. Hence, the training-based received SNR is defined as  \cite{Hassibi&Hochwald:03IT,Noh&Zoltowski&Sung&Love:14arXiv}
\begin{equation}
\mbox{SNR}_{i(l)} = \frac{\tilde{\wbf}_{i(l)}^H\left(\hat{\gbf}_{l(i)|l(i)}\hat{\gbf}_{l(i)|l(i)}^H\right)\tilde{\wbf}_{i(l)}}{\tilde{\wbf}_{i(l)}^H\left(
\Pbf_{l(i)|l(i)}+\gamma^{-1}\Ibf\right)\tilde{\wbf}_{i(l)}},\label{eq:SNR}
\end{equation}
where $\gamma :=\sigma_d^2/\sigma^2$, since $\Pbf_{l(i)|l(i)}=\Ebb\{ \Delta \gbf_{l(i)} \Delta \gbf_{l(i)}^H\}$.
The optimal beamforming vector that maximizes $\mbox{SNR}_{i(l)}$ is given by solving a generalized eigenvalue problem. In general, a closed-form solution to a generalized eigenvalue problem is not available. However, since the rank of $\hat{\gbf}_{l(i)|l(i)}\hat{\gbf}_{l(i)|l(i)}^H$ in the numerator of the RHS of \eqref{eq:SNR} is one, one can easily solve the problem in this case, and the optimal beamforming vector $\tilde{\wbf}_{i(l)}^\star$ and the corresponding optimal $\mbox{SNR}_{i(l)}^\star$ are given by
\begin{align}
\tilde{\wbf}_{i(l)}^\star&=  \left(\Pbf_{l(i)|l(i)}+\gamma^{-1}\Ibf\right)^{-1}\hat{\gbf}_{l(i)|l(i)},\\
\mbox{SNR}_{i(l)}^\star&= \hat{\gbf}_{l(i)|l(i)}^H\left(\Pbf_{l(i)|l(i)}+\gamma^{-1}\Ibf\right)^{-1}\hat{\gbf}_{l(i)|l(i)}.\label{eq:optimalSNR}
\end{align}
Note that the optimal received SNR is the same for all data symbols $i=lT+T_t+m$, $m=1,\cdots,T_d$ of each block. Hence, we shall use the notation $\mbox{SNR}_{l}^\star$ for $\mbox{SNR}_{i(l)}^\star$. Also, note from \eqref{eq:optimalSNR} that the optimal SNR is a function of symbol SNR $\gamma$, the  error covariance matrix $\Pbf_{l(i)|l(i)}$ and the  channel estimate $\hat{\gbf}_{l(i)|l(i)}$. Hence, simply minimizing the trace of $\Pbf_{l(i)|l(i)}$ may not be optimal to maximize the received SNR due to the term $\hat{\gbf}_{l(i)|l(i)}$.  Using the fact that both $\Pbf_{l(i)|l(i)}$ and $\hat{\gbf}_{l(i)|l(i)}$ are functions of the pilot signal $\tilde{\Sbf}_l$, as seen in \eqref{eq:KalmanUpdate}, we can express the optimal $\mbox{SNR}_{l}^\star$ as a function of $\tilde{\Sbf}_l$, given by
\begin{align}
\mbox{SNR}_{l}^\star &= (\hat{\gbf}_{l|l-1}+\Kbf_l(\ybf_l-\tilde{\Sbf}_l^H\hat{\gbf}_{l|l-1}))^H
\biggl((\Ibf-\Kbf_l\tilde{\Sbf}_l^H)\Pbf_{l|l-1}\nonumber\\
&+\gamma^{-1}\Ibf\biggr)^{-1}
 (\hat{\gbf}_{l|l-1}+\Kbf_l(\ybf_l-\tilde{\Sbf}_l^H\hat{\gbf}_{l|l-1})).\label{eq:SNR_random}
\end{align}
Our goal is to design the sequence $\{\tilde{\Sbf}_l,l=0,1,2,\cdots\}$ of pilot  matrices  to maximize $\mbox{SNR}_{l}^\star$. However, $\mbox{SNR}_{l}^\star$ is a function of all previous pilot signal matrices  via $\Pbf_{l|l-1}$ and  $\hat{\gbf}_{l|l-1}$, and the design problem is a complicated joint problem. Thus, as in \cite{Noh&Zoltowski&Sung&Love:13ASILOMAR,Noh&Zoltowski&Sung&Love:14arXiv}, we adopt the greedy sequential approach and the design problem is explicitly formulated as follows.

\vspace{0.3em}

\begin{problem}  \label{problem:original}
Given the channel statistics information, $a$ and $\Rbf_\hbf$, and all previous pilot matrices $\{\tilde{\Sbf}_0,\tilde{\Sbf}_1,\cdots,\tilde{\Sbf}_{l-1}\}$, design $\tilde{\Sbf}_l$ such that
\begin{equation} \label{eq:problem_formulation}
\begin{array}{cl}
\mathop{\max}\limits_{\tilde{\Sbf}_l} & \Ebb\{\mbox{SNR}_{l}^\star\} \\
\text{subject to} & \mbox{Tr}(\tilde{\Sbf}_l\tilde{\Sbf}_l^H)\le\rho T_t.
\end{array}
\end{equation}
Here, the expectation in \eqref{eq:problem_formulation} is to average out the randomness in the random vector $\ybf_l$.
\end{problem}

\section{The Proposed Design Method}
\label{sec:algorithm}

To solve Problem \ref{problem:original}, we begin with the following proposition.

\vspace{0.3em}

\begin{proposition}\label{Prop_1}
The  pilot design problem \eqref{eq:problem_formulation} is equivalent to the following optimization problem:
\begin{equation}
\begin{array}{cl}
\mathop{\min}\limits_{\tilde{\Sbf}_l}     & \mbox{Tr}\left(\Abf_l(\Bbf_l+\tilde{\Sbf}_l\tilde{\Sbf}_l^H)^{-1}\right)\\
\text{subject to} & \mbox{Tr}(\tilde{\Sbf}_l\tilde{\Sbf}_l^H)\le\rho T_t,
\end{array}
\label{eq:problem}
\end{equation}
where $
\Abf_l=\gamma\hat{\gbf}_{l|l-1}\hat{\gbf}_{l|l-1}^H+\gamma\Pbf _{l|l-1}+\Ibf$  and $\Bbf_l=\gamma\Ibf+\Pbf_{l|l-1}^{-1}$. Note that $\Abf_l$ and $\Bbf_l$ are not functions of the design variable $\tilde{\Sbf}_l$.
\end{proposition}

{\em Proof:} From \eqref{eq:optimalSNR} the average received SNR, $\Ebb\{\mbox{SNR}_{l}^\star\}$, with the optimal beamforming vector $\wbf_{i(l)}^\star$  can be expressed as
\begin{equation}
\Ebb\{\mbox{SNR}_{l}^\star\}= \mbox{Tr}\left[\left(\Pbf_{l|l}+\gamma^{-1}\Ibf\right)^{-1}\Ebb\{\hat{\gbf}_{l|l}\hat{\gbf}_{l|l}^H\}\right].\label{eq:SNR_star}
\end{equation}
 Since $\hat{\gbf}_{l|l}$ is a Gaussian random vector with mean $\hat{\gbf}_{l|l-1}$ and covariance matrix $\Qbf_l$ given by
\begin{align}
\Qbf_l=&\Pbf_{l|l-1}\tilde{\Sbf}_l(\Ibf+\tilde{\Sbf}_l^H\Pbf_{l|l-1}\tilde{\Sbf}_l)^{-1}\tilde{\Sbf}_l^H\Pbf_{l|l-1}\nonumber\\
      =&\Pbf_{l|l-1}\tilde{\Sbf}_l\tilde{\Sbf}_l^H(\Pbf_{l|l-1}^{-1}+\tilde{\Sbf}_l\tilde{\Sbf}_l^H)^{-1},
\end{align}
where the second equality holds by the matrix inversion lemma,   $\Ebb\{\hat{\gbf}_{l|l}\hat{\gbf}_{l|l}^H\}$ is given by
\begin{equation}
\Ebb\{\hat{\gbf}_{l|l}\hat{\gbf}_{l|l}^H\}=\hat{\gbf}_{l|l-1}\hat{\gbf}_{l|l-1}^H+\Pbf_{l|l-1}\tilde{\Sbf}_l\tilde{\Sbf}_l^H(\Pbf_{l|l-1}^{-1}+\tilde{\Sbf}_l\tilde{\Sbf}_l^H)^{-1}.\label{eq:hl_function}
\end{equation}
The error covariance matrix $\Pbf_{l|l}$ is expressed as
\begin{align}
\Pbf_{l|l} &= \Pbf_{l|l-1}-\Kbf_l\tilde{\Sbf}_l\Pbf_{l|l-1}\nonumber\\
           &= \Pbf_{l|l-1}-\Pbf_{l|l-1}\tilde{\Sbf}_l\tilde{\Sbf}_l^H(\Pbf_{l|l-1}^{-1}+\tilde{\Sbf}_i\tilde{\Sbf}_i^H)^{-1}.\label{eq:Pl_function}
\end{align}
Substituting  \eqref{eq:hl_function} and \eqref{eq:Pl_function} to \eqref{eq:SNR_star}, we have
\begin{align}
&\mbox{Tr}\biggl[\left(\Pbf_{l|l-1}-\Pbf_{l|l-1}\tilde{\Sbf}_l\tilde{\Sbf}_l^H(\Pbf_{l|l-1}^{-1}+\tilde{\Sbf}_l\tilde{\Sbf}_l^H)^{-1}+\gamma^{-1}\Ibf\right)^{-1}\nonumber\\
&\cdot\left(\hat{\gbf}_{l|l-1}\hat{\gbf}_{l|l-1}^H+\Pbf_{l|l-1}\tilde{\Sbf}_l\tilde{\Sbf}_l^H(\Pbf_{l|l-1}^{-1}+\tilde{\Sbf}_l\tilde{\Sbf}_l^H)^{-1}\right)\biggr] \nonumber\\
=&\mbox{Tr}\biggl[\left(\left(\Pbf_{l|l-1}+\gamma^{-1}\Ibf\right)(\Pbf_{l|l-1}^{-1}+\tilde{\Sbf}_l\tilde{\Sbf}_l^H)-\Pbf_{l|l-1}\tilde{\Sbf}_l\tilde{\Sbf}_l^H\right)^{-1}\nonumber\\
\cdot&\left(\hat{\gbf}_{l|l-1}\hat{\gbf}_{l|l-1}^H\Pbf_{l|l-1}^{-1}+(\hat{\gbf}_{l|l-1}\hat{\gbf}_{l|l-1}^H+\Pbf_{l|l-1})\tilde{\Sbf}_l\tilde{\Sbf}_l^H\right)\biggr]\nonumber\\
=&\mbox{Tr}\biggl[\gamma\left(\gamma\Ibf+\Pbf_{l|l-1}^{-1}+\tilde{\Sbf}_l\tilde{\Sbf}_l^H\right)^{-1}\left\{\hat{\gbf}_{l|l-1}\hat{\gbf}_{l|l-1}^H\Pbf_{l|l-1}^{-1} \right. \nonumber\\
&+(\hat{\gbf}_{l|l-1}\hat{\gbf}_{l|l-1}^H+\Pbf_{l|l-1})(\gamma\Ibf+\Pbf_{l|l-1}^{-1}+\tilde{\Sbf}_l\tilde{\Sbf}_l^H) \nonumber
\end{align}
\begin{align}
&\left.-(\hat{\gbf}_{l|l-1}\hat{\gbf}_{l|l-1}^H+\Pbf_{l|l-1})(\gamma\Ibf+\Pbf_{l|l-1}^{-1})\right\}\biggr] \nonumber\\
=&\gamma\mbox{Tr}\left[\hat{\gbf}_{l|l-1}\hat{\gbf}_{l|l-1}^H+\Pbf_{l|l-1}^{-1}\right]-\gamma\mbox{Tr}\left[(\gamma
\hat{\gbf}_{l|l-1}\hat{\gbf}_{l|l-1}^H\right.\nonumber\\
     &\left.+\gamma\Pbf_{l|l-1}+\Ibf)(\gamma\Ibf+\Pbf_{l|l-1}^{-1}+\tilde{\Sbf}_l\tilde{\Sbf}_l^H)^{-1}\right]. \label{eq:SNR_last}
\end{align}
Here, we used $\mbox{Tr}(\Abf\Bbf)=\mbox{Tr}(\Bbf\Abf)$ and $(\Abf\Bbf)^{-1}=\Bbf^{-1}\Abf^{-1}$. Since the first term of the RHS of \eqref{eq:SNR_last} is independent of $\tilde{\Sbf}_l$ and  the second term of the RHS of \eqref{eq:SNR_last} is $\mbox{Tr}(\Abf_l(\Bbf_l+\tilde{\Sbf}_l\tilde{\Sbf}_l^H)^{-1})$ with $\Abf_l$ and $\Bbf_l$ defined in the proposition, the problem \eqref{eq:problem_formulation} is equivalent to the problem \eqref{eq:problem}. $\hfill\blacksquare$

Note that the problem \eqref{eq:problem} is not a convex optimization problem. To tackle the problem \eqref{eq:problem}, we use the semi-definite relaxation (SDR) technique \cite{Luo&Ma&So&Ye&Zhang}. First,  introducing a new variable $\Xbf_l:=\tilde{\Sbf}_l\tilde{\Sbf}_l^H$, we change the optimization problem \eqref{eq:problem} as
\begin{equation}
\begin{array}{cl}
\mathop{\min}\limits_{\Xbf_l}     & \mbox{Tr}\left(\Abf_l(\Bbf_l+\Xbf_l)^{-1}\right)\\
\text{subject to} & \mbox{Tr}(\Xbf_l)\le\rho T_t,\\
                  & \Xbf_l \succeq \mathbf{0},\\
                  & \mbox{rank}(\Xbf_l) \le T_t.
\end{array}
\label{eq:problem_SDR_rankconstraint}
\end{equation}
Then,  dropping the rank constraint in the problem \eqref{eq:problem_SDR_rankconstraint}, we change the problem to the following optimization problem:
\begin{equation}
\begin{array}{cl}
\mathop{\min}\limits_{\Xbf_l}     & \mbox{Tr}\left(\Abf_l(\Bbf_l+\Xbf_l)^{-1}\right)\\
\text{subject to} & \mbox{Tr}(\Xbf_l)\le\rho T_t,\\
                  & \Xbf_l \succeq \mathbf{0}.
\end{array}
\label{eq:problem_SDR}
\end{equation}
Since $\Abf_l$ and $\Bbf_l$ are positive-definite matrices, the problem \eqref{eq:problem_SDR} is a convex optimization problem and can be solved by a standard convex optimization solver. To obtain a solution matrix $\tilde{\Sbf}_l^\star$ of size $R_c \times T_t$ from the solution $\Xbf_l^\star$ of \eqref{eq:problem_SDR}, we use a randomization technique. That is, we generate $T_t$ i.i.d. random vectors according to the distribution
$\Cc\Nc(\mathbf{0},\Xbf_l^\star)$. After the generation of these random vectors, we stack the vectors to make a $R_c \times T_t$ matrix $\tilde{\Sbf}_l^\star$. Since $\Abf_l$ and $\Bbf_l$ can be obtained by the standard Kalman recursion, only solving the problem  \eqref{eq:problem_SDR} and applying the randomization technique are additionally necessary to design the received-SNR-optimized pilot  sequence.

\subsection{The Block I.I.D. Channel Case}
\label{subsec:blockiid}

The block i.i.d. channel case \cite{Kotecha&Sayeed:04SP} is a
special case of  the model \eqref{eq:channel_evoluation} or
\eqref{eq:channel_evoluation2} with $a=0$. Under this model, the
Kalman recursion \eqref{eq:KalmanUpdate} is still valid although the
recursion does not propagate, i.e.,
$\hat{\gbf}_{l|l-1}={\mathbf{0}}$ and $\Pbf_{l|l-1}=\Lambdabf$ for
every $l$. Hence, Proposition \ref{Prop_1} is valid under the
block i.i.d. channel model. In this case, $\Pbf_{l|l-1}=\Lambdabf$
is a diagonal matrix and thus, the matrices $\Abf_l$ and $\Bbf_l$
in Proposition \ref{Prop_1} are diagonal. In this case, the
optimization problem \eqref{eq:problem} can be solved efficiently
without solving \eqref{eq:problem_SDR_rankconstraint} based on the
following proposition.

\vspace{0.3em}

\begin{proposition}\label{Thm_1}
There exists an optimal solution to the problem \eqref{eq:problem}
in the form of $\tilde{\Sbf}_l^\star  =\Pibf\Dbf$, where $\Pibf$ is a $R_c \times R_c$
permutation matrix and $\Dbf$ is a $R_c \times T_t$ ``diagonal''
matrix in the form of
\begin{equation}
\Dbf = \left[ \begin{array}{ccc|c}
\delta_1 & 0 & 0&\\
0&\ddots &0 & \Obf\\
0& 0 & \delta_{T_t} &\\
\end{array}  \right]^T, ~~\delta_i \ge 0~~\forall~i,
\end{equation}
when $\Abf_l$ and $\Bbf_l$ are diagonal matrices.
\end{proposition}

{\em Proof:} The proof is similar to that of \cite[Theorem
3]{Kotecha&Sayeed:04SP}. Since $\Abf_l$ is a positive definite
matrix, the objective function of the problem \eqref{eq:problem}
can be rewritten as
\begin{equation}  \label{eq:Prop2Obj}
\mbox{Tr}\left((\Abf_l^{-\frac{1}{2}}\Bbf_l\Abf_l^{-\frac{H}{2}}+\Abf_l^{-\frac{1}{2}}\tilde{\Sbf}_l\tilde{\Sbf}_l^H\Abf_l^{-\frac{H}{2}})^{-1}\right),
\end{equation}
where $\Abf_l = \Abf_l^{1/2}\Abf_l^{H/2}$. Let $\Cbf_l:=
\Abf_l^{-\frac{1}{2}}\Bbf_l\Abf_l^{-\frac{H}{2}}+\Abf_l^{-\frac{1}{2}}\tilde{\Sbf}_l\tilde{\Sbf}_l^H\Abf_l^{-\frac{H}{2}}$,
$\lambdabf(\Cbf_l):=[\lambda_1(\Cbf_l),\cdots,\lambda_{R_c}(\Cbf_l)]^T$
and $\dbf(\Cbf_l):=[\Cbf_{l}(1,1),\cdots,\Cbf_{l}(R_c,R_c)]^T$.
Then, the objective function \eqref{eq:Prop2Obj} can be rewritten
as
$f(\lambdabf(\Cbf_l)):=\sum_{i=1}^N\frac{1}{\lambda_i(\Cbf_l)}$,
since the trace of a matrix is the sum of its eigenvalues. It is
shown in \cite[Theorem 3]{Kotecha&Sayeed:04SP} that
$f(\lambdabf(\Cbf_l))$ is lower bounded by $f(\dbf(\Cbf_l))$, i.e.
$f(\lambdabf(\Cbf_l))\ge f(\dbf(\Cbf_l))$, based on the Schur
convexity of $f(\cdot)$. This lower bound  can be achieved when
$\Cbf_l$ is a diagonal matrix. To make $\Cbf_l$ a diagonal matrix,
$\Xbf_l=\tilde{\Sbf}_l\tilde{\Sbf}_l^H$ should be a diagonal
matrix, since $\Abf_l$ and $\Bbf_l$ are diagonal matrices.
Therefore, the minimum value of the objective function can be
achieved when $\tilde{\Sbf}_l\tilde{\Sbf}_l^H$ is a diagonal
matrix.  By decomposing the $R_c\times R_c$ diagonal matrix
$\Xbf_l=\tilde{\Sbf}_l\tilde{\Sbf}_l^H$ of rank less than or equal
to $T_t$, we have a solution to \eqref{eq:problem} in the form of
$\tilde{\Sbf}_l=\Pibf\Dbf$. (The locations of the non-zero
elements of $\Xbf_l$ determine $\Pibf$.) $\hfill\blacksquare$

Using Proposition \ref{Thm_1}, the  Lagrange multiplier technique
and the fact that $\Abf_l=\gamma(\Bbf_l-\gamma\Ibf)^{-1}+\Ibf$, we
obtain the optimal  diagonal elements $\{x_i\}$ of
$\Xbf_l=\tilde{\Sbf}_l\tilde{\Sbf}_l^H$ given by
\begin{align}
x_i&=\max\left(-\Bbf_{l}(i,i)+\sqrt{\frac{\Bbf_l(i,i)}{\nu(\Bbf_l(i,i)-\gamma)}},~0\right) \label{eq:block_algorithm}\\
&= \max\left(-\gamma - \frac{1}{\lambda_i(\Rbf_\hbf)} + \sqrt{\frac{\gamma \lambda_i(\Rbf_\hbf) + 1}{\nu}},~0 \right).
\end{align}
Since the object function in \eqref{eq:problem}
can be rewritten as $\sum_{i=1}^{R_c}
\frac{\Abf_l(i,i)}{\Bbf_{l}(i,i)+x_i}$ and the term
$\frac{\Abf_l(i,i)}{\Bbf_{l}(i,i)+x_i}$ is a monotone increasing
function of $\Bbf_l(i,i)$, the indices with the smallest $T_t$
$\Bbf_l(i,i)$ values should be selected for possibly non-zero
$T_t$ $x_i$'s. Let this index set be denoted by ${\mathcal{I}}$.
Then, the Lagrange multiplier $\nu$ is obtained to satisfy the
power constraint $\sum_{i\in {\mathcal{I}}}x_i=\rho T_t$ by the
bisection method. The proposed index selection here corresponds to
selecting the $T_t$ dominant eigen-directions of $\Rbf_\hbf$ since
$\Bbf_l=\gamma \Ibf + \Pbf_{l|l-1}^{-1}=\gamma\Ibf +
\Lambdabf^{-1}$. Interestingly, this index selection method
coincides with the result in \cite{Kotecha&Sayeed:04SP} minimizing
the channel estimation MSE. (The channel estimation MSE
minimizing problem is equivalent to \eqref{eq:problem} with
redefined $\Abf_l:=\Ibf$ and $\Bbf_l := \Lambdabf^{-1}$.) In both
received SNR maximization and channel estimation MSE
minimization, the $T_t$ dominant channel eigen-directions should
be used for pilot patterns, but the power allocation is a bit
different.

\vspace{0.3em}
\begin{remark}
By Proposition \ref{Thm_1}, in MISO systems with the block i.i.d.
channel model, a received-SNR-optimal pilot signal is given by
$\Sbf_l = \Ubf \Pibf \Dbf$. Hence, there is no need to mix
multiple channel  eigen-directions at a symbol time to improve the
performance. At each symbol time, it is sufficient to use one
column of $\Ubf$. On the other hand, in the block-correlated channel
case ($a\ne 0$), the optimal solution $\Xbf$ to
\eqref{eq:problem_SDR_rankconstraint} is not diagonal in general
and thus, mixing multiple channel eigen-directions at a symbol
time can improve the received SNR performance.
\end{remark}

\section{Numerical Result}
\label{sec:result}

In this section, we provide some numerical results to evaluate our
pilot  design method.  We set 2 GHz carrier frequency, $T_s=100\mu
s$ symbol duration, block size $T=10$ with three training symbols per block
($T_t=3$), and the pedestrian mobile speed $v=3km/h~(a=0.9997)$.
(The temporal fading coefficient $a$ is given by $a=J_0(2\pi f_d
T_s T)$ by Jakes' model \cite{Jakes:book}, where $f_d$ is the
maximum doppler frequency and $J_0$ is the 0-th order Bessel function.) For the channel spatial correlation
matrix $\Rbf_\hbf$, we consider the exponential correlation model
given by $\Rbf_\hbf(i,j)=r^{2|i-j|}$ with $r=0.9$.

\begin{figure}[!hbtp]
\centering
\includegraphics[width=9.5cm]{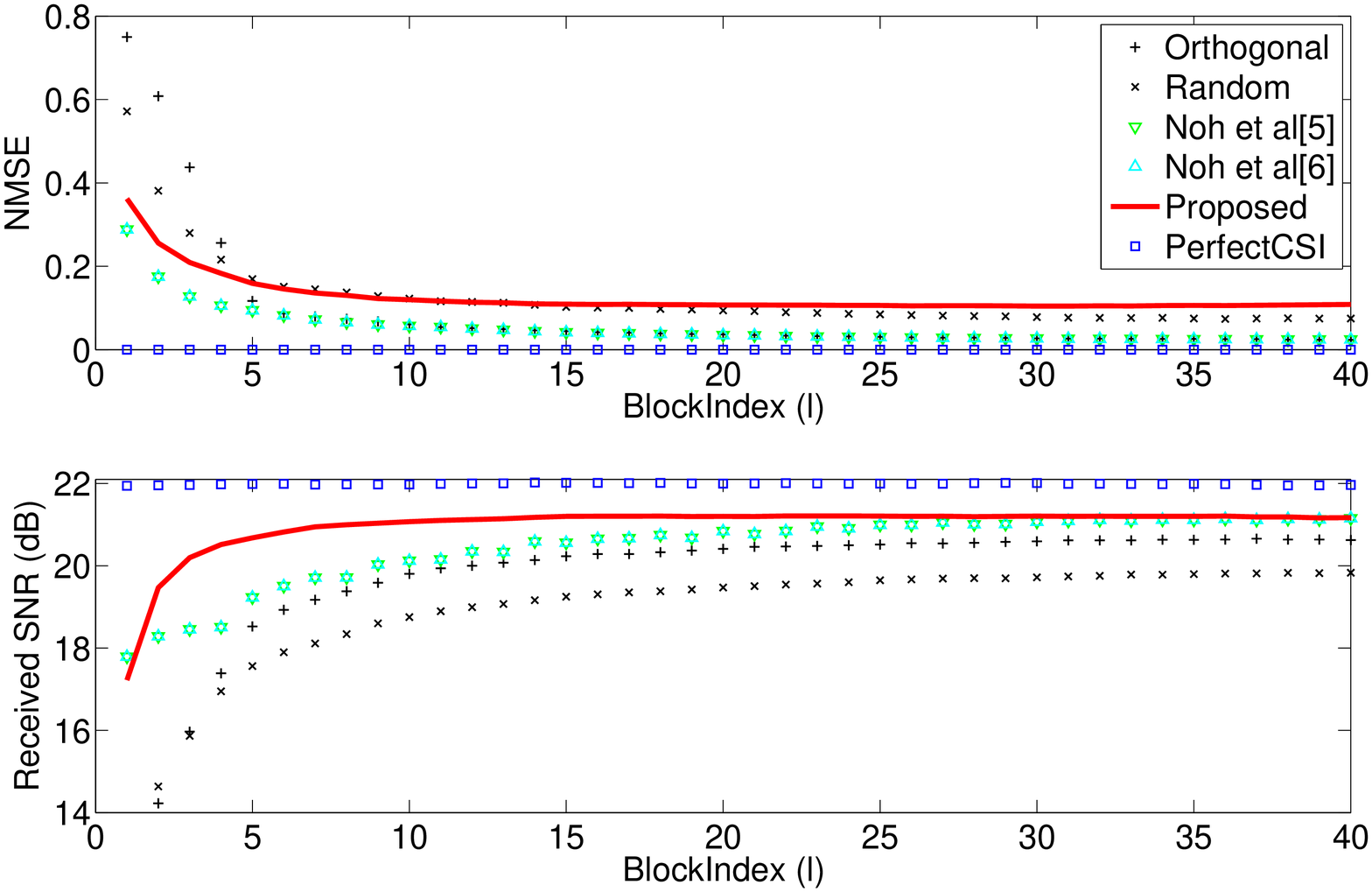}\\
\caption{NMSE and received SNR versus block index $l$:  $N=16$,
$T_t=3$, $\rho/\sigma^2=10dB$, $\gamma=10dB$, and $v=3km/h$}
\label{fig:3km_10dB}
\end{figure}

Fig. \ref{fig:3km_10dB} shows the performance of the proposed pilot design, when $\rho/\sigma^2=\gamma=$ 10dB and $N_t=16$. The
normalized MSE (NMSE) is defined as
$\frac{\|\gbf_l-\hat{\gbf}_{l|l}\|^2}{\|\gbf_l\|^2}$. The result
is averaged over 100 random realizations of the channel process
with length 40 blocks. For comparison, we consider orthogonal and
random beam patterns for $N_t=16$. In addition, we consider  the
pilot design algorithms minimizing the channel estimation MSE in
\cite{Noh&Zoltowski&Sung&Love:14arXiv,Noh&Zoltowski&Sung&Love:14ICASSP}.
It is seen that the proposed method noticeably  outperforms other
methods in terms of received SNR and especially yields quick convergence at
the early stage of channel learning, although its MSE performance
is worse than the methods in
\cite{Noh&Zoltowski&Sung&Love:14arXiv,Noh&Zoltowski&Sung&Love:14ICASSP}.
Although the result is not shown here  due to space limitation, it
is observed in the block i.i.d. channel case that the proposed
pilot design method in Section \ref{subsec:blockiid} yields
slightly better performance  than the
method in \cite{Kotecha&Sayeed:04SP} in terms of received SNR.

\section{Conclusion}
\label{sec:conclusion}

In this paper, we have considered the pilot signal design
for massive MIMO systems to maximize the received SNR under the
block Gauss-Markov and block i.i.d. channel models. We have shown that
the proposed design method yields noticeably better performance in
terms of received SNR than channel estimation MSE-based methods.
Furthermore, we have shown that using the $T_t$ dominant
eigen-vectors of the channel covariance matrix without mixing as
the pilot signal provides an optimal solution even for received
SNR maximization under the block i.i.d. channel model. The extension to the MIMO case is left as future work.
%





\end{document}

%% file: macros.tex
\def\psfancypar#1#2{\begingroup\def\par{\endgraf\endgroup\lineskiplimit=0pt}
               \setbox2=\hbox{\large\sc #2}
               \newdimen\tmpht \tmpht \ht2 \advance\tmpht by \baselineskip
               \font\hhuge=Times-Bold at \tmpht
               \setbox1=\hbox{{\hhuge #1}}
               \count7=\tmpht \count8=\ht1
               \divide\count8 by 1000 \divide\count7 by \count8 
               \tmpht=.001\tmpht\multiply\tmpht by \count7 
               \font\hhuge=Times-Bold at \tmpht
               \setbox1=\hbox{{\hhuge #1}}
               \noindent
                \hangindent1.05\wd1
               \hangafter=-2 {\hskip-\hangindent
               \lower1\ht1\hbox{\raise1.0\ht2\copy1}%
                \kern-0\wd1}\copy2\lineskiplimit=-1000pt}

\newcommand{\E}{\mbox{{\rm E}}}

\def\boxit#1{\vbox{\hrule\hbox{\vrule\kern3pt
        \vbox{\kern3pt#1\kern3pt}\kern3pt\vrule}\hrule}}

\def\reals{ { {\rm  I \kern-0.15em R }  } }
\def\complex{ {\,{{\rm C} \kern-0.50em \raise0.20ex {  |}}\, }}

\def\lambdabf{\hbox{\boldmath$\lambda$\unboldmath}}
\def\mubf{\hbox{\boldmath$\mu$\unboldmath}}

\def\Sigmabf{\hbox{$\bf \Sigma$}}

\def\Lambdabf{\mbox{$ \bf \Lambda $}}

\def\Pibf{{\bf \Pi}}

\def\bbf{{\bf b}}

\def\dbf{{\bf d}}
\def\ebf{{\bf e}}

\def\gbf{{\bf g}}
\def\hbf{{\bf h}}

\def\nbf{{\bf n}}

\def\sbf{{\bf s}}

\def\wbf{{\bf w}}
\def\xbf{{\bf x}}
\def\ybf{{\bf y}}

\def\xbf{{\bf x}}
\def\ybf{{\bf y}}
\def\Abf{{\bf A}}
\def\Bbf{{\bf B}}
\def\Cbf{{\bf C}}
\def\Dbf{{\bf D}}

\def\Ibf{{\bf I}}

\def\Kbf{{\bf K}}

\def\Obf{{\bf O}}
\def\Pbf{{\bf P}}
\def\Qbf{{\bf Q}}
\def\Rbf{{\bf R}}
\def\Sbf{{\bf S}}

\def\Ubf{{\bf U}}

\def\Xbf{{\bf X}}

\def\Cc{{\cal C}}

\def\Lc{{\cal L}}

\def\Nc{{\cal N}}

\def\be{\vskip .3cm \begin{equation}}
\def\ee{\end{equation} \vskip .4cm \noindent}
\newcommand{\R}{\mbox{$\hat {\bf R}_{N}$}}

\def\Rxx{\Rbf_{\ssstyle X\kern-.1em X}}

\let\ssstyle=\scriptscriptstyle

\def\Kout{\setbox1=\hbox{\Huge\bf K}\hbox to
1.05\wd1{\hspace{.05\wd1}
}}
\def\Sout{\setbox1=\hbox{\Huge\bf S}\hbox to 1.05\wd1{\hspace{.05\wd1}
}}

%% file: labelfig.tex
  \ifx\LabelFigloaded\MYundefined\relax
  \else
   \fi

  \def\LabelFigloaded{\relax}

  \chardef\LabelFigCatAt\the\catcode`\@
  \catcode`\@=11

 \let\LabelFigwlog@ld\wlog
 \def\wlog#1{\relax}

 \ifx\\\MYundefined@
    \let\\\relax
 \fi

  \def\ms@g{\immediate\write16}

 \def\N@wif{\csname newif\endcsname }
 \def\Temp@ {\N@wif\ifIN@}
 \ifx\INN@\MYundefined@
    \else \let\Temp@\relax
 \fi
 \Temp@

  \def\IN@{\expandafter\INN@\expandafter}
  \long\def\INN@0#1@#2@{\long\def\NI@##1#1##2##3\ENDNI@
    {\ifx\m@rker##2\IN@false\else\IN@true\fi}%
     \expandafter\NI@#2@@#1\m@rker\ENDNI@}
  \def\m@rker{\m@@rker}
 
  \newtoks\Initialtoks@  \newtoks\Terminaltoks@
  \def\SPLIT@{\expandafter\SPLITT@\expandafter}
  \def\SPLITT@0#1@#2@{\def\TTILPS@##1#1##2@{%
     \Initialtoks@{##1}\Terminaltoks@{##2}}\expandafter\TTILPS@#2@}

 \def\Shifted@@#1#2#3{\setbox0=\hbox{#3}%
   \raise -\dp0\vbox {\kern-#2%
       \hbox {\kern#1\unhbox0\kern-#1}%
           \kern#2}}

 \newcount\gridcount
 \newbox\auxGridbox@ \newbox\hGridbox@ \newbox\vGridbox@
 \newbox\Labelbox@ \newbox\auxLabelbox@
 \newbox\Coordinatebox@
 \newtoks\Labeltoks@
 \newdimen\Wdd@ \newdimen\Htt@
 \newdimen\Wddd@ \newdimen\Httt@
 
 \def\Wr@{\immediate\write16}

 \newdimen\GL@wd
 \def\GridLineWidth#1{\GL@wd=#1}

 \def\gobble#1{}
 \def\EdgeErr@{\Wr@{}%
      \Wr@{\string\Edges\space argument
      1, 10, 100 or 1000 please\string!}%
      }

 \newcount\Edgect@

 \def\Sweepup#1\endSweepup{}

 \def\SetEdges@{%
    \edef\Zr@@s{\expandafter\gobble\number\Edgect@\empty}%
        \count255=0\Zr@@s\relax
        \ifnum\count255=\z@\else\EdgeErr@\show\tailtest\fi
        \count255=1\Zr@@s\relax
        \ifnum\count255=\Edgect@\relax\else\EdgeErr@\show\leadtest\fi
    \EdgGl@b\edef\Zr@s{\expandafter\gobble\Zr@@s\empty}
    \ifnum\Edgect@>\@ne\relax\EdgGl@b\let\L@Dc\empty
        \else\EdgGl@b\edef\L@Dc{\string.}\fi
    \ifnum\Edgect@>\@ne\relax
        \EdgGl@b\edef\Edgescale@##1{\divide##1 by \Edgect@}%
        \else\EdgGl@b\edef\Edgescale@##1{}\fi
    }

 \def\Edges#1{\Edgect@=#1\relax
     \let\EdgGl@b\global \SetEdges@}

 \Edges{1}

 \def\hhrule{\hrule height \GL@wd\vskip-.\GL@wd}

 \def\hRule@{%
   \advance\gridcount -2%
   \vfil\hhrule\vfil
   \llap{\smash{\raise -2.5pt
     \hbox{\L@Dc\number\gridcount\Zr@s\kern2pt}}}%
   \hhrule
   }

\def\vvrule{\vrule width \GL@wd \kern-\GL@wd}

 \def\vRule@{\advance\gridcount 2%
   \hfil\vvrule\hfil
   \setbox\auxGridbox@=\vbox to 0pt
      {\vskip \Htt@\vskip 2pt
        \hbox to 0pt{\hss\L@Dc\number\gridcount\Zr@s\hss}\vss}%
      \wd\auxGridbox@=0pt \box\auxGridbox@
   \vvrule
   }

 \def\PlaceGrid@@{\gridcount=10 
  \setbox\hGridbox@=\hbox{%
        \hbox{%
             \hskip-.4pt\vrule
             \vbox to \Htt@{%
               \offinterlineskip\parindent=\z@\relax
               \hbox to \Wdd@{\hfil}
               \hRule@\hRule@\hRule@\hRule@
               \vfil\hhrule\vfil}%
             \vrule\hskip-.4pt}
    }%
  \gridcount=0%
  \setbox\vGridbox@=\hbox{%
      \vbox{\offinterlineskip\parindent=0pt\hsize=0pt
         \vskip-.4pt\hrule%
         \hbox to \Wdd@{%
                 \vtop to \Htt@{\vfil}%
                 \vRule@\vRule@\vRule@\vRule@
                 \hfil\vvrule\hfil}%
         \hrule\vskip-.4pt}}%
  \wd\hGridbox@=0pt\ht\hGridbox@=0pt
  \wd\vGridbox@=0pt\ht\vGridbox@=0pt
  \hbox{\box\hGridbox@\box\vGridbox@}%
  }

 \def\LabelsGlobal{\def\LabGl@b{\global}}
 \def\LabelsLocal{\def\LabGl@b{}}
 \LabelsGlobal 

 \def\SetLabels#1\endSetLabels{%
   \LabGl@b\Labeltoks@={#1()\\}%
   }

 \LabGl@b\Labeltoks@={()\\}

 \def\ShowGrid{\LabGl@b\let\PlaceGrid@\PlaceGrid@@}
 \def\HideGrid{\LabGl@b\let\PlaceGrid@\relax}
 \def\Grids{\ShowGrid\LabGl@b\let\GridSwitch@\ShowGrid}
 \def\noGrids{\HideGrid\LabGl@b\let\GridSwitch@\HideGrid}

 \noGrids

 \def\bAdjust@@{%
     \setbox\auxLabelbox@=\hbox{\raise \dp\auxLabelbox@
            \box\auxLabelbox@}}
 \def\bAdjust@{\let\vAdjust@\bAdjust@@}

 \def\eAdjust@@{\dimen0=-.5\ht\auxLabelbox@
     \advance\dimen0 by .5\dp\auxLabelbox@
     \setbox\auxLabelbox@=
            \hbox{\raise\dimen0\box\auxLabelbox@}}
 \def\eAdjust@{\let\vAdjust@\eAdjust@@}

 \def\tAdjust@@{%
     \setbox\auxLabelbox@=\hbox{\raise-\ht\auxLabelbox@
            \box\auxLabelbox@}}
 \def\tAdjust@{\let\vAdjust@\tAdjust@@}

 \let\vAdjust@\relax

 \def\lAdjust@{\let\hAdjust@\rlap}
 \def\rAdjust@{\let\hAdjust@\llap}

 \let\hAdjust@\relax\let\vAdjust@\relax

 \def\FetchLabel@#1(#2)#3\\{%
     \IN@0#2@@\ifIN@
        \setbox0=\hbox{\ignorespaces#1#3\unskip}%
        \ifdim\wd0>0pt
           \ms@g{}%
           \ms@g{ !!! Bad label(s)? !!!}%
           \message{ #1(#2)#3}%
        \fi
        \def\LabelMole@##1\endFetchLabel@{%
            \IN@0()\\@##1@%
            \ifIN@\def\Temp@{\FetchLabel@##1\endFetchLabel@}%
            \else\def\Temp@{}%
            \fi
            \Temp@
           }%
     \else
       \ignorespaces#1\unskip
       \setbox\auxLabelbox@=%
         \hbox to 0pt{\hss\ignorespaces\hAdjust@
          {\ignorespaces#3\unskip}\hss}%
       \vAdjust@
       \let\hAdjust@\relax\let\vAdjust@\relax
       \AugmentLabelBox@@{#2}%
       \ht\Labelbox@=0pt\dp\Labelbox@=0pt
       \let\LabelMole@\FetchLabel@%
     \fi\LabelMole@}

 \newtoks\XYSep@ 
 \def\SetXYSeparator#1{%
     \IN@0#1@@\ifIN@\XYSep@{*}%
     \else
     \XYSep@{#1}%
     \fi
     }

 \SetXYSeparator*

 \def\AugmentLabelBox@@#1{%
     \IN@0\the\XYSep@ @#1@\ifIN@
       \SPLIT@0\the\XYSep@ @#1@%
       \setbox\Labelbox@=\hbox to 0pt{%
         \unhbox\Labelbox@
         \Shifted@@{\the\Initialtoks@\Wddd@}%
         {\the\Terminaltoks@\Httt@}%
         {\box\auxLabelbox@}}%
     \else
         \ms@g{}%
         \ms@g{ !!! Bad insertion point. !!!}%
         \message{ (#1\ this point was rejected.)}%
     \fi
    }

 \def\FetchOption@#1[#2]#3\endFetchOption@{%
    \def\temp{#1}
    \ifx\temp\empty
       \Edgect@=#2\relax
       \let\EdgGl@b\relax
       \SetEdges@
       \Cleaner@#3%
    \fi}

 \def\Cleaner@#1[@]{\Labeltoks@{#1}}
     
 \def\PlaceLabels@@{\mathsurround=0pt
     \def\Cr@{\\}%
     \let\L\lAdjust@\let\R\rAdjust@
     \let\B\bAdjust@\let\E\eAdjust@\let\T\tAdjust@
     \expandafter\FetchOption@\the\Labeltoks@[@]\endFetchOption@
     \Wddd@=\Wdd@ \Edgescale@\Wddd@ 
     \Httt@=\Htt@ \Edgescale@\Httt@
     \expandafter\FetchLabel@\the\Labeltoks@\endFetchLabel@
     \box\Labelbox@
     }%

 \let \PlaceLabels@\PlaceLabels@@

 \def\AffixLabels#1{\setbox\Coordinatebox@=\hbox{#1}%
      \Wdd@=\wd\Coordinatebox@ \Htt@=\ht\Coordinatebox@
      \advance\Htt@ \dp\Coordinatebox@
      \hbox{\copy\Coordinatebox@\kern-\Wdd@ 
           \Shifted@@{0pt}{-\dp\Coordinatebox@}%
           {\PlaceLabels@\PlaceGrid@}%
           \kern\Wdd@}%
      \GridSwitch@ 
      \LabGl@b\Labeltoks@{()\\}%
      }
 
   \let\wlog\LabelFigwlog@ld   
   \catcode`\@=\LabelFigCatAt  

                                By

              Raymond S\'eroul <A18645@FRCCSC21.BITNET>
                                and 
              Laurent Siebenmann <lcs@topo.math.u-psud.fr>
    
              VERSIONS: July 1991, Oct 1991, Jan 1992, July 1992

INTRODUCTION

      This labelling package is intended for TeX users who
rely on non-TeX sources for for their graphics inserts.  It
provides means for adding TeX labels to such inserts with a
minimum of fuss. 

       For most labels, TeX users have in the past found it
reasonably convenient to rely on non-TeX sources. Typical
occasions when an inescapable need for TeX labels seemed to
arise are

 (a) when the graphics program lacks certain exotic or complex
mathematical symbols

 (b) when the very highest typographical quality is wanted for the
labels

 (c) when labels included with the graphics fail to print, 
 and you cannot figure out why (cf. boxedeps.doc).  The labels
 provided by labelfig.tex are 100

       Since this package first appeared, many users, who in the
past scarcely dreamed of using TeX labels, have come to use
nothing but.  So it is now appropriate to add

Intoxication Warning:  TeX labels may be addictive and expensive. 

     If you have a fast preview you may disagree, and even find
that this package provides an agreeable paste-up environment; see
extra applications at end.

     Note to publishers: It is possible and convenient to ultimately
export the TeX labels produced by labelfig.tex to become an integral
part of the EPS file. This is often desired by a publisher who typically
uses an "upmarket" graphics or page layout program, with which the
staff is skilled in perfecting figures.  See Appendix I for
a recipe.

     The authors are grateful to Patrick Ion of Math Reviews for
helpful comments and encouragement.

BASIC INSTRUCTIONS

    After reading in the macro file using

\input labelfig.tex
preview or proof your figure with a coordinate grid printed on
top, by typing the following:

    \ShowGrid  
    \AffixLabels{<the graphics insertion>}

Here <the graphics insertion> is what you would type to insert
the graphics object alone without the grid.  This must provide
for the space around it. For example <the graphics insertion>
might well be \BoxedEPSF{MyFigure scaled 700} using the
boxedeps.tex macro package (from same source); this provides a
TeX box containing the encapsulated PostScript insert specified by
the file MyFigure. \AffixLabels{...} provides the grid (supposing
\ShowGrid is present) and later, once you have specified labels
using the grid, it will "tack on" the labels.

     The grid is a sort of (usually elongated) checkerboard of
ten rows and ten columns and its (internal) partitions are by
default numbered  .1, ... ,.9  both horizontally (X-coordinate
running left to right) and vertically (Y-coordinate running bottom
to top).  Thus the points enclosed by the grid correspond to the
points of the unit square in the cartesian "X-Y" plane, the lower
left corner corresponding to the origin (0,0).  By extrapolation,
the full page corresponds to a larger rectangle in the plane.

     These coordinates serve to position labels as follows.
Before the \AffixLabels{...} command type label specifications:

  \SetLabels
   (<X-coordinate>*<Y-coordinate>) <first label> \\
   .
   .
   .
   (<X-coordinate>*<Y-coordinate>)  <last label> \\
  \endSetLabels

Each row specifies one label and is terminated by \\.  In each
row, the position indicator comes first; it is written as a
standard cartesian point except that the X- and Y- coordinates
are separated by * rather than a comma because TeX allows a
comma as decimal point. There are no dimension units to specify
as the unit is the grid itself.

     By default, this cartesian point specifies where the middle
of the baseline of the label will be located.  However if you precede
the point by \L [or \R] the left [or right] edge of the baseline will
be located there. Similarly you may also precede the point by \T, \E,
or \B to vertically align the top equator or bottom of the label box
at the specified point.  This gives nine standard positions of
the label with respect to the insertion point --- corresponding to
the eight principle points of the compas and the center

                     \L\T     \T      \R\T

                     \L\E     \E      \R\E

                     \L\B     \B      \R\B

But this neglects the default "baseline" level of TeX,
giving potentially three more positions

                     \L    <no tag>   \R

For text, the baseline level is often the preferred. Its relation to
the others is variable. It will often coincide with the bottom level,
as happens for "X".  But it is often distinct, as for "g", in which
case you have in all 12 distinct positions rather than 9.

     It is convenient to think of this specification of label
position as attaching the label by a thumb-tack to the coordinate
grid. There are up to twelve positions of the thumb-tack on the
label, while the position of the thumb-tack on the coordinate grid is
arbitrary.  Normally, one choses the position of the thumb-tack on
the label to be the one that is the closest to the item being
labeled.  There are good reasons for this "rule of thumb":

   (a)  It facilitates correct positioning at first try.

   (b)  If the scale of the figure must be altered after labels
have been affixed, the labels have a good chance of remaining well
positioned.

   (c)  The visible grid need not extend beyond the "bounding box"
for the figure, because the best preferred position is always
(at least almost) within the bounding box .

The second reason is particularly important. Indeed it often
happens that scale has to be altered after labelling begins, in
order to either provide space for the labels, or to adjust
proportions between the labels and the figure.  (The size of labels
is unaffected by scaling.)

     Here is an artificial but self-contained test which uses
TeX rules to make a graphics object.

TEST

    Do not skip this!

\input labelfig

 \def\FrameIt#1{\hbox{\vrule$\vcenter {\hrule\kern3pt%
             \hbox {\kern3pt #1\kern3pt}%
               \kern3pt\hrule}$\relax\vrule}}

 \def\Caption#1#2{\FrameIt{%
       \vtop {\hsize=#1\relax \parindent=0pt
         \leftskip=0pt \rightskip=0pt plus15pt
         \parfillskip=0pt
         \lineskip=1pt\baselineskip=0pt
         #2}}}

 \def\FirstQuadrant{\hbox to 100pt{\vrule\vbox to 100pt{%
        \hbox to 100pt{\hfil}\vfil\hrule}\hss}}

  \SetLabels
    \R(.5*.2) $\zeta\,\cdot$\\
    (.9*-.10) $\xi$\\
    \R(-.03*.9) $\eta$\\
    \T(.5*.9) \Caption{70pt}{%
          \it The norm of
          $g(\xi+i\eta)$ is indicated on
          contours of this invisible surface.}\\
  \endSetLabels

  \AffixLabels{\FirstQuadrant}

  \end

  Note that the coordinates to use for labels are indicated on the
edges of the grid (when visible) corresponding to the conventional
x- and y- axes of the Cartesian plane. By default the grid is
1-by-1. However, by the command \Edges{100}, you can change this
to 100-by-100 and many users find this alternative most
convenient. Place the command \Edges{...} in your style file (or
header) since its effect is is global. Other possible edge values
are 10 and 1000.

  If you use the command \Edges{...} at all, do so with care.  For
if you accidentally delete an \Edges{...} command your labels will
abruptly be badly misplaced and may logically but mysteriously
generate "dimension too big" errors under TeX and "off page" errors
under your driver.  

  You can dictate the edgescale for an individual figure by giving
the scale in brackets immediately after \SetLabels.  Thus, to
import into an article using say \Edge{100} a figure labelled using
another edgescale, say the original 1-by-1 default, you can use
\SetLabels[1]...\endSetLabels.

GETTING IT DOWN PAT

     Complicated labeling deserves the same respect as
complicated mathematics.  Do not expect it to come out perfect the
first time!  What is needed in either case is a mechanism to
repeatedly typeset troublesome pieces.

     One mechanism is always available.  One does complicated
labelling in a separate "test" file involving just the figure being
labelled;  a texpert will know how to \dump TeX's current state as
a temporary format that restarts rapidly at each retry.  Usually,
one then pastes the completed labelled figure back into the main
TeX file, but, of course, one can also \input it as an auxiliary
file.

     If you do not have a TeXpert at handy, here is a first
approximation to an efficient setup. By deletions reduce a copy
of your article to just a few lines before and after the figure.
Now label the figure, and finally, copy and paste the labelled
figure to the original article. Then copy the next figure to label
into this testbed and repeat. The TeXpert can improve the  speed
at which TeX starts up, by compiling a format specifically for
your article; just one caution: best NOT include in the format
ephemeral details of setup like \Set<mydriver>ArtSpecials (from
boxedeps.tex because this reads  figure dimensions which you may
change during your work session.

     An improved mechanism to repeatedly typeset troublesome
pieces is now available on the Macintosh; it is called LinoTeX;
see the same ftp sources.  It could be set up on many types
of computer.

     Before using labelfig.tex to attach labels to a graphics
object inserted using boxedeps.tex or BoxedArt.tex, make it a
firm rule to carefully adjust the bounding box using the trimming
commands of these packages, and also at least tentatively scale
and position the object. Beware of changing the grid inadvertently
after the labels have been positioned.  For example, correcting
the bounding box of a PostScript graphics object can foul up the
labels by changing the coordinate grid to which the labels are
attached. This is particularly true for the trimming  commands of
boxedeps.tex and BoxedArt.tex. However, as noted already, change
of scale is much less disruptive, and modest adjustments should be
well tolerated.

     Sometimes the labels protrude so far from the bounding box
of a figure that the figure has to be repositioned.  Best do this
by ad hoc spacing, say using \hglue and \vglue; altering the
bounding box would create a vicious circle.

     Remember that you are responsible for preventing labels
from overlapping. You are responsible for all label typography
including size and style. A label is really just about anything
that can be put in a TeX box. Note that spaces at the beginning
and end of labels will normally be suppressed; if you really want
them you must protect them with TeX braces.

     This package temporarily sets the \mathsurround parameter
of TeX to zero  while the labels are being affixed. This is done
because nonzero \mathsurround space would influence the position
of left and right aligned labels; then, when a texpert or printer
modifies mathsurround, diagram labeling might be disastrously
altered. There is a small price to pay involving labels that are
formatted as caption boxes including mathematics: you  may want or
need to specify an explicit mathsurround space within the caption
box; it will not influence anything outside.

     Those hostile to the use of * as separator between
the X and Y coordinates of label insertion points, are free to
impose another using \SetXYSeparator{<the new separator>}.  
Americans may prefer "," to "*" since they never use a 
comma as a decimal point; on the other hand, * may be more visible.

APPENDIX (I)  MERGING labelfig.tex LABELS INTO AN EPSF GRAPHICS OBJECT.

     As promised in the introduction, here is a recipe useful for
publishers. It works at least on Macintosh and at least for vectorized
graphics and Adobe type1 fonts.  (There is surely a similar recipe for
PCs under MSWindows.)

 (a)  Use boxedeps.tex utility to integrate the figure given by the eps
file, "x.eps" say, with a visible frame around it.  See
\ShowDisplacementBoxes command in boxedeps.tex.  To get precise results
automatically it is important to use the \Trim... commands of
boxedeps.tex making the "DisplacementBox" neatly fit the figure.

 (b)  Use the TeX printer driver and LaserWriter (versions >= 8.1.1) to
export to an EPSF the DVI page containing the integrated, labelled
figure. You now have an EPS file  "xx.eps"  that contains too much, and at
the wrong scale, and at wrong position.

 (c)  Convert the EPSF to an Adode Illustrator format EPSF using
the shareware utility called epsConvert by Sam Weiss
1993-- (currently $25).

 (d)  In Illustrator (or a compatible program), group the labels and the
"DisplacementBox"; copy them to the clipboard and paste them into "x.ps".
This step requires that all the label fonts be "visible to the Macintosh.

 (e)  Translate and scale the pasted group consisting of the labels plus
the "DisplacementBox" so as to make the "DisplacementBox" the bounding
box of (labelless) figure represented by "x.eps".  At this point the
labels will be correctly placed on the figure "x.eps".

 (f)  Ungroup and delete the "DisplacementBox".  The result is the
desired single EPS file, "x+.eps" say, It contains the original figure
plus its labels.  

     Using grouping and ungrouping appropriately in "x+.eps", a
publisher's staff can very efficiently improve label positions etc.

APPENDIX II)  SOME EXOTIC APPLICATIONS

     The grid of labelfig.tex is analogous to a light-table in
classical page makeup with wax or latex glue.  In principle, you
can use it to compose any page from its indivisible parts.  This
even has some of the artisanal charm of classical paste-up
provided you have a fast screen preview to make the process
"interactive".

     In practice labelfig.tex is a tool for nonstandard jobs.
Here are a few going beyond the labelling already discussed.

(I)  GRAPHICS INTEGRATION.

     This is accomplished by treating the imported graphics
objects as labels.  The underlying graphics object is then
typically an empty  \vbox to <dimension>{\vfill} in a TeX
\midinsert...\endinsert construction.  A label line
might be of the form

   (.1*.1) \\

The exact form of the special command varies from driver to
driver.  However, in the case of encapsulated PostScript graphics
(EPSF norm), by relying on boxedeps.tex, one can have the
following standard syntax (independant of driver  (see
boxedeps.doc for details.
  
  (.1*.1) \BoxedEPSF{MyFigure scaled <scale in mils>}\\

This may be slow since it requires TeX to read the PostScript
file to read bounding box using many complex macros.  So you
may want to try

  (.1*.1) \EPSFSpecial{MyFigure}{<scale in mils>}\\

which is fast and driver independant, but it squashes the
bounding box, normally to its lower left corner.

     Similarly for graphics of the Macintosh PICT norm ---
using BoxedArt.tex (same sources) in place of boxedeps.tex.

     This approach to integration is to be recommended when
one is assembling a composite graphics object.

 (II)  COMMUTATIVE DIAGRAM ENHANCEMENT

     Commutative diagrams or arrays of mathematical objects
connected by arrows of various sorts are common in mathematics.
The mathematical objects require the use of TeX.  Recently TeX
acquired a good collection of arrows of all slopes --- that of
LamSTeX --- plus pwerful macros to build the diagrams.

     However, even the LamSTeX collection is often
inadequate; it lacks for example double shafted arrows, dotted
arrows and curved arrows. Fortunately it is possible to produce
such arrows on an individual basis using sophisticated graphics
programs such as Illustrator and AldusFreehand (both serving
the EPSF norm) or using Metafont (with its public domain norm).
Since the creation of each new arrow is a work of love, you
probably want to limit the number of arrows by using LamSTeX
for most arrows. The 40K commutative diagram module of LamSTeX
has been adapted to work with AmSTeX and a copy may be posted
with LabelFig and related files. Unfortunately no one has yet
offered a version that works with Plain TeX or LaTeX.

       Suffice it here to say that when the exotic arrow has
been somehow imported into TeX, labelfig.tex treats it as a
label that one affixes to the commutative diagram.  Two other
steps will be treated in separate notes, namely the matter of
extracting the dimension specifications for the arrow and the
construction of the arrow --- for these steps are far from
unique and often depend intimately on your computer environment. 
Notes for the Macintosh-Textures-Illustrator combination are
found in the file ExoticArrows.doc.

 (III) NESTING 

Ingenuity pays off in exploiting labelfig.tex. One can
mix graphics and typography quite freely.  labelfig.tex is good
for freeform or overlapping arrangements, while boxedeps.tex (or
BoxedArt.tex) is best for regimented non-overlapping
arrangements --- and the two can be combined.

     The default behavior of labelfig.tex is not ideal 
for nesting objects, because to prevent trouble for beginners
the register for labels is globally cleared when \AffixLabels
concludes.  But there are switches available

      \LabelsGlobal      \LabelsLocal

which change this.  To understand this, extend the above test 
by something like:

 \LabelsLocal

 \SetLabels
    (.5*.5) AAA\\
 \endSetLabels

 {
 \SetLabels
    (.5*.5) ZZZ\\
 \endSetLabels
   \AffixLabels{\FirstQuadrant}
 }

   \AffixLabels{\FirstQuadrant}

     There are however potential pitfalls.  Neither
labelfig.tex nor boxedeps.tex has been tested under extreme
conditions. Problems may occur if their procedures are
indiscriminately nested. For boxedeps.tex (not labelfig.tex)
there is a precise cause for worry, namely many of its
variables are "global", which means that TeX braces will not
provide the protection one might expect.

COMMAND SUMMARY FOR labelfig.tex

  Here [...] means optional (one or zero)
       [...]* means any number of such constructs

  \SetLabels
    [[<P>](<X><Sep><Y>) <label> \\]*
  \endSetLabels
  \ShowGrid  
  \AffixLabels{<the figure>}

   --- <P> is tack position, one of eleven or empty
              order irrelevant

                   \L\T      \T      \R\T

                   \L\E      \E      \R\E

                     \L               \R

                   \L\B      \B      \R\B

   --- (<X><Sep><Y>) insertion point;
  <Sep> is separator, = * by default;
  \SetXYSeparator{<Sep>} changes it.
   <X> and <Y> are real numbers

  --- <label> a label to attach 

  --- <the figure> the figure to label 

  \GlobalLabels (default)     
  \LocalLabels  setting for nested constructs.

 \Grids makes ALL grids appear; \HideGrid then makes just next disappear.
 \noGrids returns to default.  The commands are always global.

 \GridLineWidth{<dimension>} adjusts width of grid lines. Default is very
small, to give "hairline" effect. If your grid lines are missing try
setting \GridLineWidth{1pt}.

 \Edges#1 globally changes the edge size of all grids to the numerical 
value #1, which must be 1, 10, 100, or 1000.  The default is 1.

VERSION HISTORY.
 --- Jan 1993: \Edges#1 and [??] option after \SetLabels
 --- July 1992: \Grids, \noGrids, \HideGrid;
       Gridlines become hairlines; \GridLineWidth{<dimension>}.
 --- Oct 1991, Jan 1992: \SetXYSeparator{<Sep>},  \LabelsGlobal,
       \LabelsLocal.
 --- July 1991: first release

Address for bugs and other feedback:

        Raymond S\'eroul
        IREM and Lab. de Typographie Informatise
        Univ. Rene Descartes
        Strasbourg

    Tel 33-88-41-63-45
    Email:  A18645@FRCCSC21.BITNET

        Laurent Siebenmann
        Mathematique, Bat. 425,
        Univ de Paris-Sud,
        91405-Orsay,
        France

    Tel 33-1-6941-7949; 
    Email: lcs@topo.math.u-psud.fr